\documentclass[aps,prl,twocolumn,showpacs,amsmath,amssymb,footinbib]{revtex4}
\usepackage{graphicx}
\usepackage{amsmath}
\usepackage{enumitem}
\usepackage{natbib}
\usepackage{float}
\usepackage{verbatim}
\usepackage{color}

\begin{document}

\title{How many quasiparticles can be in a superconductor?}
\author{Anton Bespalov, Manuel Houzet, Julia S. Meyer}
\affiliation{Univ.~Grenoble Alpes, INAC-PHELIQS, F-38000 Grenoble, France}
\affiliation{CEA, INAC-PHELIQS, F-38000 Grenoble, France}
\author{Yuli~V.~Nazarov}
\affiliation{Kavli Institute of NanoScience, Delft University of Technology, Lorentzweg 1, NL-2628 CJ, Delft, The Netherlands}

\begin{abstract}
Experimentally and mysteriously, the concentration of quasiparticles in a gapped superconductor at low temperatures always by far exceeds its equilibrium value. We study the dynamics of localized quasiparticles in superconductors with a spatially fluctuating gap edge. 
The competition between phonon-induced quasiparticle recombination and generation by a weak non-equilibrium agent results in an upper bound for the concentration that explains the mystery.
\end{abstract}

\pacs{74.40.Gh, 	%Nonequilibrium superconductivity
74.62.En, 			%Effects of disorder
74.81.-g 			%Inhomogeneous superconductors and superconducting systems, including electronic inhomogeneities
}

\maketitle

Na\"{i}vely, the superconducting gap $\Delta$ should ensure an exponentially small quasiparticle concentration at low temperatures. However, various experiments indicate that a long-lived, non-equilibrium quasiparticle population persists in the superconductor~\cite{Martinis2009, Lenander2011, deVisser2011,Rajauria2012, Riste2013, Wenner2013, LevensonFalk2014}. The quasiparticle poisoning, whereby an unwanted quasiparticle is trapped in a bound state, is an important factor harming the ideal operation of superconducting devices~\cite{Catelani2011}. Unwanted quasiparticles also forbid tempting perspectives to use Majorana states in superconductors for topologically protected quantum computing~\cite{Fu2009, vanHeck2011, Rainis2012}.
The poisoning rates have been quantified~\cite{BretheauRPL, Bretheau2013, Bretheau2013b,vanWoerkom2015,Higginbotham2015} and much experimental work is directed on protection from poisoning, with important advances in this direction~\cite{Peltonen,Wang2014,Vool2014,Plourde2014}. 
The non-equilibrium quasiparticles are produced by some non-equilibrium agent, which is most likely related to the absorption of electromagnetic irradiation from the high-temperature environment~\cite{PekolaHekking} and/or electromagnetic fields applied to the setup in the course of its measurement and operation. Surprisingly, the efforts to reduce the intensity of this non-equilibrium agent are not entirely satisfying: the experiments give a substantial residual quasiparticle concentration, even if all efforts are performed~\cite{Pekola,Klapwijk}.
 
In this Letter, we study the dynamics of the annihilation of quasiparticles localized at the spatial fluctuations of the gap edge. Importantly, we find that the average distance between the quasiparticles depends only {\it logarithmically} on the intensity of the non-equilibrium agent. 
In simple terms, the exponential dependence of the annihilation rate on the distance between the two quasiparticles results in the
quasiparticle concentration
\begin{equation}
\label{eq:c}
c = \frac{C_p}{({4\pi}/{3}) r^{3}} \quad {\rm with} \quad  \frac r {r_c} \approx  \ln\left(\frac{\bar{\Gamma}}{A r^6_c}\right),
\end{equation}
valid at small $A\ll \bar \Gamma/r_c^6.$
[A more accurate estimate for $r$ is given by Eq.~\eqref{eq:accurater}.]
Here, $r_c$ is the relevant radius of the localized quasiparticle state to be estimated in detail below: for practical circumstances, it exceeds the superconducting coherence length $\xi_0$ by not more than an order of magnitude. Furthermore, $A$ is the rate of non-equilibrium generation of quasiparticles per unit volume, and $\bar{\Gamma}$ is a material constant characterizing the inelastic quasiparticle relaxation due to electron-phonon interaction.
The packing coefficient, $C_p \approx   0.605\pm0.008 $, can be derived from a simple {\it bursting bubbles} model outlined below. 
Equation~\eqref{eq:c} explains both the substantial concentration that is observed, as well as the inefficiency of the efforts to reduce it. 

Let us outline the derivation of the above relations. The relevant quasiparticles have energies close to the gap edge, and they annihilate by emitting a phonon with energy $\sim 2\Delta$. Assuming the ``dirty'' limit, $\ell \ll  v_F/\Delta$, where $v_F$ is the Fermi velocity, and the phonon wavelength not exceeding the mean free path $\ell$, we derive a remarkably simple relation for the annihilation rate of two quasiparticles~\cite{SM},
\begin{equation}
\Gamma_{12} = \bar{\Gamma} \int d{\bf r}\, p_{1}({\bf r}) p_2({\bf r}).
\label{eq:rate}
\end{equation}
Here $p_{1,2}({\bf r})$ are the normalized probability densities to find the quasiparticles 1,2 at position ${\bf r}$. 
Furthermore we find~\cite{SM}
$\bar{\Gamma}= 24\, \gamma(\Delta)/(\nu_0 \Delta)$, where $\nu_0$ is the normal-metal density of states and $\gamma(\Delta)$ is the normal-metal electron-phonon relaxation rate at energy $\Delta$. For aluminium, this yields $\bar{\Gamma} \simeq  40\,$s$^{-1}\mu{\rm m}^3$.
Equation~\eqref{eq:rate} is valid for localized as well as for delocalized states. 

For large enough quasiparticle concentrations (in particular for delocalized states), one can neglect the correlations in their positions. In that case, a simple mean-field calculation~\cite{book-kinetics} shows that the balance between generation of non-equilibrium quasiparticles and their annihilation, $A = \bar{\Gamma} c^2$, results in the non-equilibrium concentration $c = (A/\bar{\Gamma})^{1/2}$. In this regime, a generation rate $A\approx 4 \times10^3\,$s$^{-1}\mu{\rm m}^{-3}$ would thus result in $c\approx10\,\mu{\rm m}^{-3}$. However, the annihilation itself reduces the probability for quasiparticles to be close to each other. Therefore it boosts the non-equilibrium concentration. This effect is most pronounced if the quasiparticles are in localized states and do not move.

The description of the quasiparticle bound states is provided in Ref.~\cite{LarkinOvchinnikov} and has been recently revisited~\cite{SkvortsovFeigelman} in the context of strongly disordered superconductors. The main results can be summarized as follows. The short-range fluctuations of the pairing potential shift the gap edge, $E_g=\Delta-\varepsilon_g$, by $\varepsilon_g \ll \Delta$ and smooth the density of delocalized states 
%at energy $E=E_g-\varepsilon$ near $\Delta$ 
on the same scale $\varepsilon_g$.
%\begin{equation}
%\nu(\varepsilon)=\nu_0\sqrt{\frac {-3 \Delta\varepsilon}{2\varepsilon_g^2}}\quad {\rm at}\quad 0<-\varepsilon\ll \varepsilon_g.
%\end{equation}
The long-range fluctuations of the pairing potential generate a tail of localized states at energies $E<E_g$. As the typical extent of these states is much larger than the correlation length of the pairing potential fluctuations, the latter can be regarded as point-correlated, $\langle\langle\Delta({\bf r})\Delta({\bf r}')\rangle\rangle = (\delta\Delta)^2 \delta({\bf r}-{\bf r}')$. The intensity of the fluctuations is conveniently characterized by a dimensionless parameter 
%$F \simeq 0.045 \, (\delta \Delta)^2/(\Delta^2 \xi_0^3)$, where 
 $F = a_1 \, (\delta \Delta)^2/(\Delta^2 \xi_0^3)$, where 
$a_1\simeq 0.045$~\cite{numbers},
$\xi_0=\sqrt{ D/\Delta}$ is the diffusive coherence length and $ D$ is the diffusion constant in the normal metal. 
%(choose the coefficient conveniently). 
For a typical localized state with energy $E<E_g$, the energy distance from the edge, $\varepsilon=E_g-E$, is of the order of the typical fluctuation, $\delta \Delta /L^{3/2}(\varepsilon)$, on the length scale $L(\varepsilon)$ of this state. The length scale itself depends on energy, $L(\varepsilon) = \xi_0 [2\Delta/(3\varepsilon)]^{1/4}$.
%(need exact coeff here) 
From this, one derives the energy scale of the exponential tail, $\varepsilon_T =  F^{4/5}\Delta$, and the corresponding length scale, $L_T=L(\varepsilon_T)\approx 0.90\,\xi_0 /F^{1/5}$.
At $\varepsilon_T\ll\varepsilon\ll\varepsilon_g$, the density of states reaches an exponential asymptotics, 
%(fix the factor)
\begin{equation}
\nu(\varepsilon) \simeq \nu_T \left(\varepsilon/\varepsilon_T\right)^{9/8} \exp[-(\varepsilon/\varepsilon_T)^{5/4}] 
\label{eq:dos}
\end{equation} 
where $\nu_T=a_2\, \nu_0 \sqrt{\varepsilon_T\Delta/\varepsilon_g^2}$ and $a_2\simeq 0.79$~\cite{numbers}, and the most probable shape of the localized state is given by 
\begin{equation}
p_{\rm LO}({\bf r})= \frac {f\boldsymbol{(}r/L(\varepsilon)\boldsymbol{)}}{2\pi L^{3}(\varepsilon)}\quad {\rm with}\quad  f(x) \equiv \frac{\sinh x}{x \cosh^3x}. 
\label{eq:shape}
\end{equation}

\begin{figure}
\includegraphics[width = \linewidth]{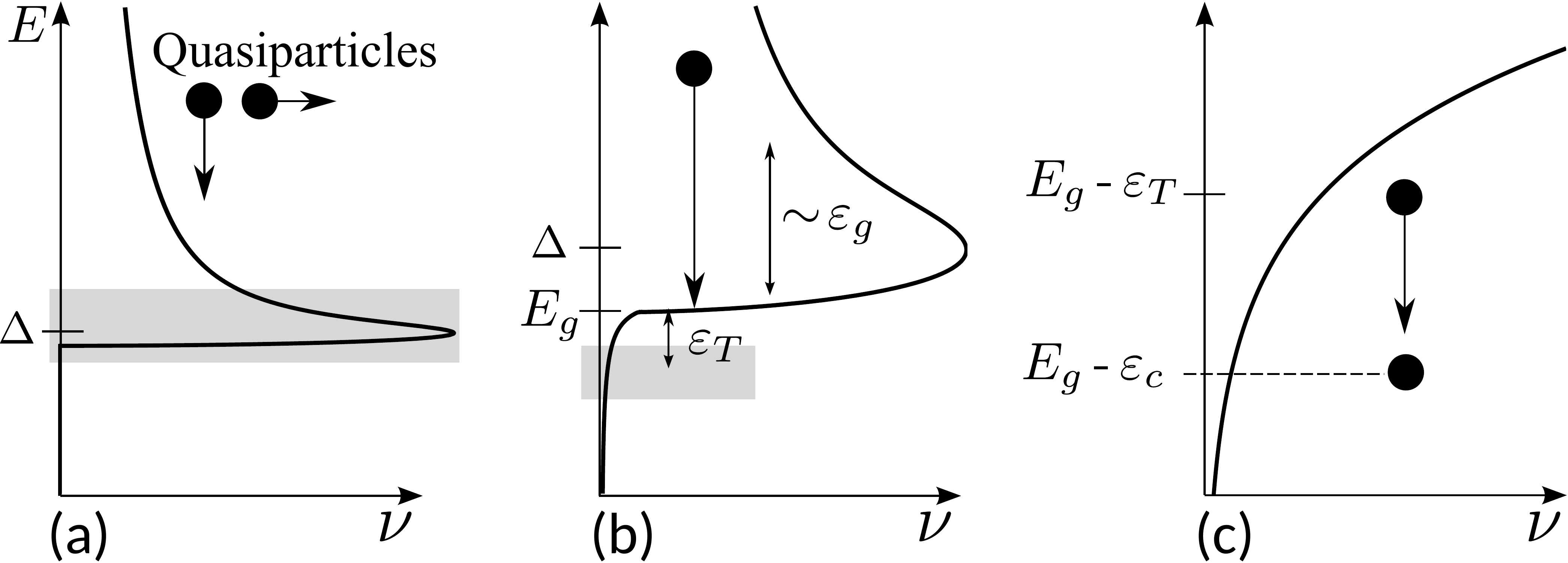}
\caption{\label{fig1}  The density of states and single quasiparticle relaxation in a superconductor. (a) The density of states is BCS-like except near the gap edge. (b) Near the gap edge, the singularity is rounded at an energy scale $\varepsilon_g$ and a tail of localized states within the gap develops at an energy scale $\varepsilon_T$. (c) The relaxation of a single quasiparticle stops at an energy scale $\varepsilon_c > \varepsilon_T$, where the localized states with lower energy no longer overlap.}
\end{figure}

Let us consider a quasiparticle generated by a non-equilibrium agent. Typically, its energy is much larger than $\Delta$. However, it loses its energy quickly due to low-energy electron-phonon  interactions before annihilating with another quasiparticle. At some stage, the quasiparticle reaches the gap edge and becomes localized at $\varepsilon \simeq \varepsilon_T$. It is important for us to understand that its relaxation does not stop here. One can estimate the number of localized states that overlap with a given state and have a lower energy, 
\begin{equation}
N(\varepsilon) \equiv L^3(\varepsilon)
\int_{\varepsilon}^{\infty} \!\!\!\!d \varepsilon' \nu(\varepsilon')
\simeq N_T(\varepsilon/\varepsilon_T)^{1/8}   \exp[-(\varepsilon/\varepsilon_T)^{5/4}] 
\end{equation}
with $N_T=(4/5) \nu_T\varepsilon_T L_T^3$.
This number is likely to be big at $\varepsilon \simeq \varepsilon_T$, where $ N(\varepsilon_T)\sim N_T\sim g\sqrt[4]{\varepsilon_T^3\Delta}/\varepsilon_g
\gg 1$~\cite{note1}. Here $g =\pi \nu_0\Delta \xi_0^3$ is 
the number of Cooper pairs in a cube of size $\xi_0$.
Thus, the quasiparticle will relax further from these states, and the relaxation stops only at a rather definite energy $\varepsilon_c$~\cite{note2} defined by  $N(\varepsilon_c) \simeq 1$, 
$\varepsilon_c \approx \varepsilon_T (\ln N_T)^{4/5}$.
Therefore we come to a rather unexpected conclusion: the quasiparticles end up their random relaxation process at a rather definite radius, $r_c \equiv L(\varepsilon_c)/2$, that is,
\begin{equation}
r_c \simeq 0.45\,  \xi_0
{(\varepsilon_T/\Delta)^{-1/4}(\ln N_T)^{-1/5}}
\end{equation} 
as illustrated in Fig.~\ref{fig1}. Taking standard parameters for Al~\cite{note3}, we expect that scale to be only slightly larger than  half the coherence length,  $\xi_0\approx  100$~nm. For instance, taking $\varepsilon_T/\Delta=10^{-2}$ and $\varepsilon_T/\Delta = 10^{-4}$, we find $r_c\approx \xi_0$ and $3\xi_0$, respectively.

%%%%%%%%%%%%
%%%%%%%%%%%%

Using these results, we can formulate a model of stochastic quasiparticle dynamics ~\cite{Chemistry}. The quasiparticles appear in random points with the rate $A$, keep their positions, and annihilate pairwise with a rate $\Gamma({\bf R})$ that is a function of their mutual distance ${\bf R}$. The rate is obtained from Eqs.~\eqref{eq:rate} and \eqref{eq:shape}. Namely,
\begin{equation}
\label{eq:rate-loc}
\Gamma({\bf R}) = \bar\Gamma\int d{\bf r} \; p_{\rm LO}({\bf r})p_{\rm LO}({\bf r}+{\bf R}) \equiv\bar{\Gamma} r_c^{-3} g(R/r_c), 
\end{equation}
where $g(2x)=(16\pi \sinh^4x)^{-1} (3+2\sinh^2x - 3 \cosh x \sinh x/x)$. 
In particular, $g(x) \simeq  1/(60\pi)$ at $x\ll1$ and $g(x) \simeq  1/(2\pi) \exp[-x]$ at $x\gg1$.

The behavior of the model is governed by a single dimensionless parameter, $A r_c^6/\bar{\Gamma}$. At large values of this parameter, the typical distance  between quasiparticles, $r$, is much smaller than $r_c$, and correlations are negligible as $\Gamma(r)\sim \bar\Gamma r_c^{-3}$ 
is constant on that length scale. In this limit, we recover the mean-field result given above, $c=(A/\bar{\Gamma})^{1/2}$,
 which is independent of $r_c$ and does not rely on tail states.
At small values of the parameter $A r_c^6/\bar{\Gamma}$, $r$ %the typical distance between quasiparticles 
is much larger than $r_c$.  In this limit, it %the typical distance 
can be estimated from the competition of the annihilation rate, $\sim \bar{\Gamma}r_c^{-3} \exp[-r/r_c]$, and the generation rate within the typical volume of a quasiparticle, $\sim A r^3$. Thus, $r \simeq r_c \ln[\bar{\Gamma}/(Ar_c^6)]$. 

\begin{figure}
\includegraphics[width = \linewidth]{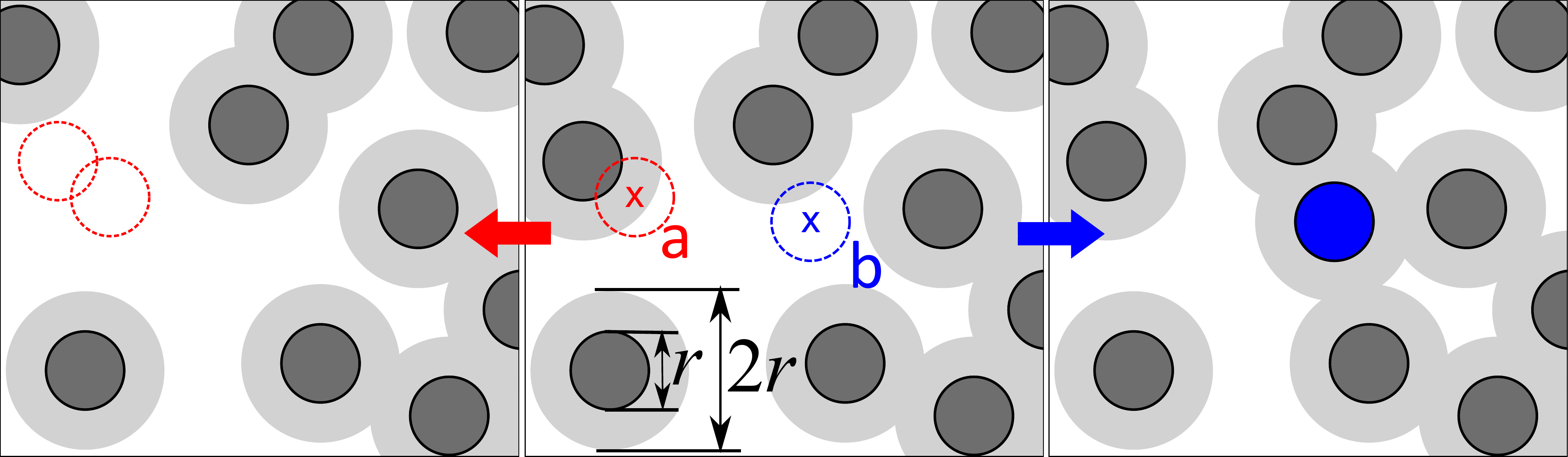}
\caption{\label{fig2} Illustration of the {\it bursting bubbles} model. Each particle is
represented by a bubble with diameter $r$ (dark gray). If a new particle
appears centered in the gray area with diameter $2r$ (case a, central panel), it
immediately annihilates with another particle (left panel). If the
particle appears in the white area (case b), it is simply added to the
system (right panel).
}
\end{figure}

Due to the exponential dependence of the annihilation rate on the typical distance,
one of the rates prevails over the other completely if the distance is changed by $\delta r\sim r_c \ll r$. This allows one to introduce a simplified model of {\it bursting bubbles},  see Fig.~\ref{fig2}. Regardless the details of $\Gamma({\bf R})$, we can consider the quasiparticles as spherical bubbles of radius $r/2$. If two bubbles overlap, the particles annihilate. This model is easily simulated: we add bubbles to the system at random points. If the added bubble does not overlap with the existing ones, the number of quasiparticles is increased by 1. If there is an overlap, two bubbles burst, decreasing the number by 1. Equilibrium is achieved when these two outcomes happen with equal probabilities. This is the case when the volume covered by spheres of radius $r$ centered around the quasiparticles equals half of the whole volume. If we rather na\"{i}vely assume that the spheres do not overlap, the volume covered is $4\pi r^3/3$ per quasiparticle, and the concentration is $c = C_p (4\pi r^3/3)^{-1}$ with $C_p=0.5$. In reality, some spheres overlap, so the simulation yields a slightly bigger packing coefficient, see Eq.~\eqref{eq:c}.

%%%%%%%%%%%%
%%%%%%%%%%%%

\begin{figure*}
\includegraphics[width =0.9  \linewidth]{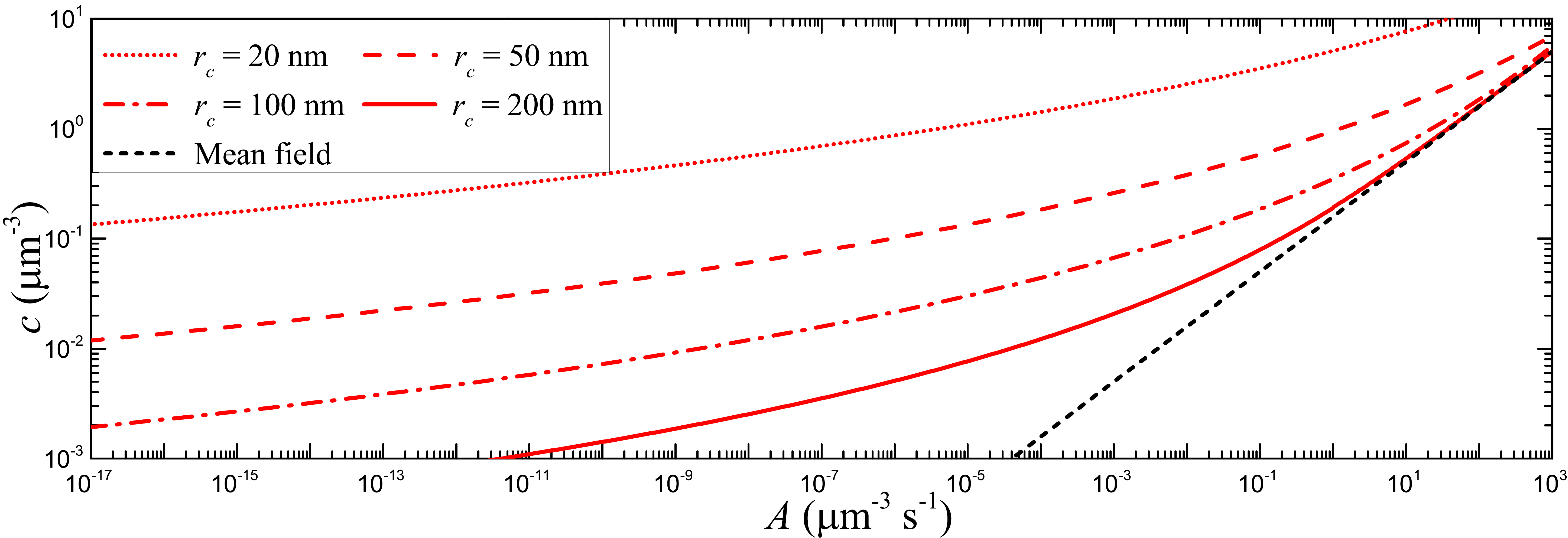}
\caption{\label{fig3}  Concentration $c$ as a function of the generation rate $A$ for quasiparticles annihilating pairwise with the rate given by Eq.~\eqref{eq:rate-loc} with $\bar{\Gamma} = 40\,$s$^{-1}\mu{\rm m}^3$, and several values of the quasiparticle localization radius $r_c$. The  line shows the mean field estimate, $c=\sqrt{A/\bar \Gamma}$, for comparison.
}
\end{figure*}

To improve upon the logarithmic estimation of $r$, we performed  simulations of the full model taking into account the details of $\Gamma({\bf R})$~\cite{SM}. The stationary concentrations are shown in Fig.~\ref{fig3}.

In the limit $\tilde{r} \equiv r/r_c \gg 1$,
the dynamics of the quasiparticle concentration is given by an evolution equation
$\dot c(t)=A- \Gamma_{\rm fit} c(t)$,
 with the effective asymptotic relaxation rate $\Gamma_{\rm fit}(r)=  4\pi/(3 C_p) b\bar\Gamma r_c^{-3}\tilde{r}^\beta e^{-\tilde{r}}$. Expressing  $c(t) =C_p/[4\pi (\tilde{r} r_c)^{3}/3]$ and introducing dimensionless time in units  of $9C_pr_c^3/(4\pi\bar \Gamma)$, this equation simplifies to 
\begin{equation}
\label{eq:fit}
\dot{\tilde{r}} = (A r_c^6/\bar \Gamma) \tilde{r}^4 - b \tilde{r}^{\beta+1} e^{-\tilde{r}}
\end{equation}
The parameters $b$ and $\beta$ can be obtained by fitting the simulation at small values of $Ar_c^6/\bar\Gamma$ with the stationary solution of Eq.~\eqref{eq:fit}  determined from 
\begin{equation}
\label{eq:accurater}
A r_c^6/\bar \Gamma = b \tilde{r}^{\beta-3} e^{-\tilde{r}} 
\end{equation}
that improves the accuracy of Eq. 1. We find
$\beta=0.41$ and $b=0.008$~\cite{SM}. 
At larger values of $Ar_c^6/\bar\Gamma$, corresponding to $\tilde r\lesssim 3.0$, the dependence of the concentration crosses over to the square-root law discussed above.

%%%%%%%%%%%%
%%%%%%%%%%%%

If the non-equilibrium agent ceases to work, $A=0$, the quasiparticles concentration relaxes very slowly. In particular, Eq.~\eqref{eq:fit} yields the estimate $\tilde{r}(t)\propto\ln(\bar\Gamma t/r_c^3)$.  Beyond the logarithmic approximation, the results of the simulation~\cite{SM} are consistent with those obtained from the stationary solution.

Using realistic values for the generation rate, we thus can give accurate estimates of the quasiparticle concentration.
 In particular, cosmic radiation is dominated by protons with energy in the GeV range and a flux of $\sim 1$~s$^{-1}$cm$^{-2}$~\cite{CosmicRays}. The stopping power of GeV protons in aluminum is $\sim 1$~MeV~cm$^{-1}$~\cite{StoppingPower}. Assuming a perfect conversion into quasiparticles of the deposited energy, we thus find a generation rate $A\sim 10^{-5}\,{\rm s}^{-1}\mu{\rm m}^{-3}$ ($\sim 1\,{\rm day}^{-1}\mu{\rm m}^{-3}$!). At $r_c\sim 0.1\,\mu$m, it yields a quasiparticle concentration $c\sim 0.01 \,\mu{\rm m}^{-3}$, which is close to the one measured in two recent experiments~\cite{Pekola,Riste2013}, where best efforts where performed in screening electromagnetic radiation.

 In the above considerations, we have assumed that the annihilation rate does not depend on the spin state of two quasiparticles. This is valid in two cases: i) the localization radius $r_c$ exceeds the spin-orbit relaxation length, which may be relevant for heavy-atom metals; ii) the spin coherence time of an isolated quasiparticle is shorter than the {(exponentially long)} timescale  $\Gamma_{{\rm fit}}$ for annihilation. 
In the opposite regimes, the quasiparticles could only annihilate if in a spin-singlet state. 

To account for the spin structure is a challenging task owing to complex quantum entanglement of the spins of the overlapping quasiparticles that survive the annihilation. As a simplifying description, we considered an extension of the bursting bubbles model in which each bubble is assigned a classical spin degree of freedom. Whenever two bubbles with opposite  spins overlap, they burst. 
The result of our simulation~\cite{SM} is an enhanced $C_p\approx {2.19\pm 0.05}$. 
When spin-flip processes are added, the concentration decreases down to $C_p\approx 0.61$ upon increasing the spin-flip rate, in agreement with the above considerations for the spinless case. 

The validity of our estimation is limited by a variety of complex factors that can influence the non-equilibrium quasiparticle dynamics in superconductors. 
In particular, we assumed immobile quasiparticles, which is valid in the limit of a vanishing temperature. At finite temperature, the quasiparticles could diffuse owing to inelastic transitions, even if they reside in localized states. This would favor their annihilation as they would come closer to each other. As a result, the estimate for the concentration given in this Letter is rather an upper bound at a given generation rate. The evaluation of the diffusion of localized quasiparticles, as well as its complex temperature dependence, would be a subject of interesting research that is needed to understand the details of their dynamics.

\begin{acknowledgments}
We thank  M.~Devoret, J.~Pekola, and F.~Portier for useful discussions. This work is supported by the Nanosciences Foundation in Grenoble, in the frame of its Chair of Excellence program, and by the ANR through the grant ANR-12-BS04-0016-03.
\end{acknowledgments}

\pagebreak

\setcounter{equation}{0}
\setcounter{figure}{0}
\setcounter{table}{0}
\setcounter{page}{1}

\onecolumngrid

\begin{center}
\textbf{\large Supplemental Material for ``How many quasiparticles can be in a superconductor?''}
\end{center}

\renewcommand{\theequation}{S\arabic{equation}}
\renewcommand{\thefigure}{S\arabic{figure}}
\renewcommand{\thesection}{S\Roman{section}}

\twocolumngrid

This Supplemental Material contains technical details on the derivation of the annihilation rate, Eq.~(2) in the main text, and on the numerical simulations, which were omitted in the main text.

\section{Annihilation rate of two quasiparticles}
\label{sec:rate}
In this Section, we derive Eq.~(2) in the main text for the annihilation rate of two quasiparticles with energies close to $\Delta$.
For this we consider a model where electrons interact with longitudinal phonons,
\begin{eqnarray}
%\begin{equation}
\hat{H}&=&\sum_{n\sigma}E_n\hat{\gamma}_{n\sigma}^\dagger \hat{\gamma}_{n\sigma}
+\sum_{{\bf q}}\omega_{q} \hat{b}_{\bf q}^\dagger \hat{b}_{\bf q}
\nonumber\\
&&+ \frac 1{\sqrt{\cal V}}\sum_{{\bf k}{\bf q}\sigma}C\sqrt{q}\,\hat{a}^\dagger_{{\bf k}+{\bf q}\sigma} \hat{a}_{{\bf k}\sigma}( \hat{b}_{\bf q}+\hat{b}_{-{\bf q}}^\dagger).
%\end{equation}
\label{Heph}
\end{eqnarray}
Here $\hat{b}_{\bf q}$ is an annihilation operator for an acoustic phonon with wavevector $\bf q$ and energy $\omega_q=v_sq$, where $v_s$ is the sound velocity, $\hat{\gamma}_{n\sigma}$ is an annihilation operator for a Bogoliubov quasiparticle with orbital label $n$, spin $\sigma=\pm$, and energy $E_n$. Furthermore, $C$ characterizes the strongly screened electron-lattice interaction, and the annihilation operator for an electron with wavevector $\bf k$ and spin $\sigma$ is
\begin{equation}
\hat{a}_{{\bf k}\sigma}=\sum_n \int  \frac {d{\bf r}}{\sqrt{\cal V}}\, e^{-i{\bf k}.{\bf r}}\left[u_n({\bf r}) \hat{\gamma}_{n\sigma}-\sigma v_n^*({\bf r}) \hat{\gamma}^\dagger_{n-\sigma}\right],
\end{equation}
where $(u_n({\bf r}),v_n({\bf r}))^T$ is the Bogoliubov-de~Gennes wavefunction associated with state $n$, and $\cal V$ is the volume of the system.
%\end{widetext}

Considering Eq.~\eqref{Heph} in the normal state, we first use the Fermi golden rule to obtain the scattering rate for a normal electron with energy $E$ at zero temperature \cite{rammer-book},
\begin{equation}
\gamma(E)=\frac{C^2E^3}{6\pi v_s^3v_F}\quad{\rm at} \quad 0<E\ll \omega_D,
\label{eq:gamma}
\end{equation}
where $\omega_D$ is the Debye frequency.
Then, in the superconducting state, we obtain the annihilation rate for two quasiparticles with opposite spins and orbital labels $n$ and $m$,
\begin{eqnarray}
\Gamma_{nm}&=&  \sum_{\bf q}\left|
\int d{\bf r}\, e^{-i{\bf q}.{\bf r}}\left[u_n({\bf r})v_m({\bf r})+v_n({\bf r}) u_m({\bf r})\right]\right|^2
\nonumber\\
&& \qquad\qquad \times  \frac {2\pi }{{\cal V}} C^2 q\,\delta(\omega_q-E_n-E_m).
\label{eq:wR}
\end{eqnarray}
As $u_n({\bf r})\approx v_n({\bf r})$ for a state with energy $E_n\approx \Delta$, Eq.~\eqref{eq:wR} simplifies to
\begin{eqnarray}
\Gamma_{nm}&=&192 \frac{v_F}{v_s} \gamma(\Delta)
\int d{\bf r} \int d{\bf x} \, \frac{\sin q_0 x}{q_0 x} 
\label{eq:wR2}
\\
&& \qquad\qquad \times u_n({\bf r+x})u_n^*({\bf r})u_m({\bf r+x}) u_m^*({\bf r}),
\nonumber
\end{eqnarray}
where $q_0=2\Delta/v_s$. In the semiclassical approximation \cite{Berry1977}, the space variables $\bf r$ and $\bf x$ describe long-range and short-range variations of the wavefunctions, respectively, such that 
\begin{equation} 
\label{eq:semiclass} 
u_n({\bf r+x})u_n({\bf r})\approx \frac{1}{2\nu_0} p_n({\bf r})
 g_N({\bf x}).
%\langle e^{-i{\bf k} \cdot {\bf x}}\rangle_{\rm F.S.}. 
\end{equation}
Here $p_n({\bf r})=2 |u_n({\bf r})|^2$ is the probability density for the quasiparticle to be at position $\bf r$, and  $g_N({\bf x})=\nu_0 [{\sin (k_F x)}/{(k_F x)}]e^{-x/(2\ell)}$, where $\nu_0$ is the normal-metal density of states, $k_F$ is the Fermi wave number,  and $\ell$ is the mean free path, is the normal-metal spectral function at the Fermi level. Assuming $k_F^{-1} \ll q_0^{-1}\ll \ell$, %where $k_F$ is the Fermi wave\change{number}, 
we insert Eq.~\eqref{eq:semiclass} into Eq.~\eqref{eq:wR2} to obtain Eq.~(2) from the main text.

To estimate the material constant $\bar{\Gamma}=24 \gamma(\Delta)/(\nu_0 \Delta)$ in aluminum, we first notice that the rate $\gamma(\Delta)$ can be related with the characteristic electron-phonon relaxation time $\tau_0$ that is introduced in Ref.~\cite{Kaplan}, $\gamma(\Delta) = (3\tau_0)^{-1} (\Delta/k_B T_c)^3$,
where $T_c$ is the superconducting critical temperature and $k_B$ is the Boltzmann constant. In aluminum, $\tau_0 \approx 400\,$ns \cite{Kaplan} and the superconducting gap at zero temperature satisfies $\Delta/k_B T_c \approx 1.76$.
Then, using $\nu_0 \approx 2\times 10^{10}\,$eV$^{-1}\mu$m$^{-3}$ and $\Delta\approx 200\,\mu$eV, we find $\bar{\Gamma}  \approx 40 \,$s$^{-1}\mu$m$^3$.

\section{Simulation of the dynamics of spinless quasiparticles with annihilation rate $\Gamma(R)$}

To determine the equilibrium concentration as a function of the quasiparticle injection rate, we performed a numerical simulation of our model with pairwise annihilating particles. The simulation volume was a box with periodic boundary conditions and size $L \times L \times L$. In each simulation step, either a quasiparticle was generated at a random position with probability $p_{\rm gen}=AL^3/(\Gamma_{\rm tot}+AL^3)$, or two quasiparticles were annihilated with probability $p_{\rm an}=1-p_{\rm gen}=\Gamma_{\rm tot}/(\Gamma_{\rm tot}+AL^3)$. %, such that $p_{\rm gen} + p_{\rm an} = 1$. 
Here $\Gamma_{\rm tot}=\sum_{i<j}\Gamma({\bf R}_i-{\bf R}_j)$, where ${\bf R}_i$ are the quasiparticle positions, and $\Gamma(\vec{R})$ is given by Eq.~(8) in the main text. For the annihilation process, a specific pair $(i,j)$ is chosen with probability $p_{ij}=\Gamma({\bf R}_i-{\bf R}_j)/\Gamma_{\rm tot}$. To acquire the equilibrium concentration in a broad range of injection rates, $A r_c^6/\bar{\Gamma} = 1.6\times 10^{-14} - 1.6 \times 10^{-1}$, the size of the box was varied in the range $L = 26r_c - 200r_c$, such that the steady-state number of particles in the box was $N = 30 - 8000$. 

\begin{figure}
\includegraphics[width = \linewidth]{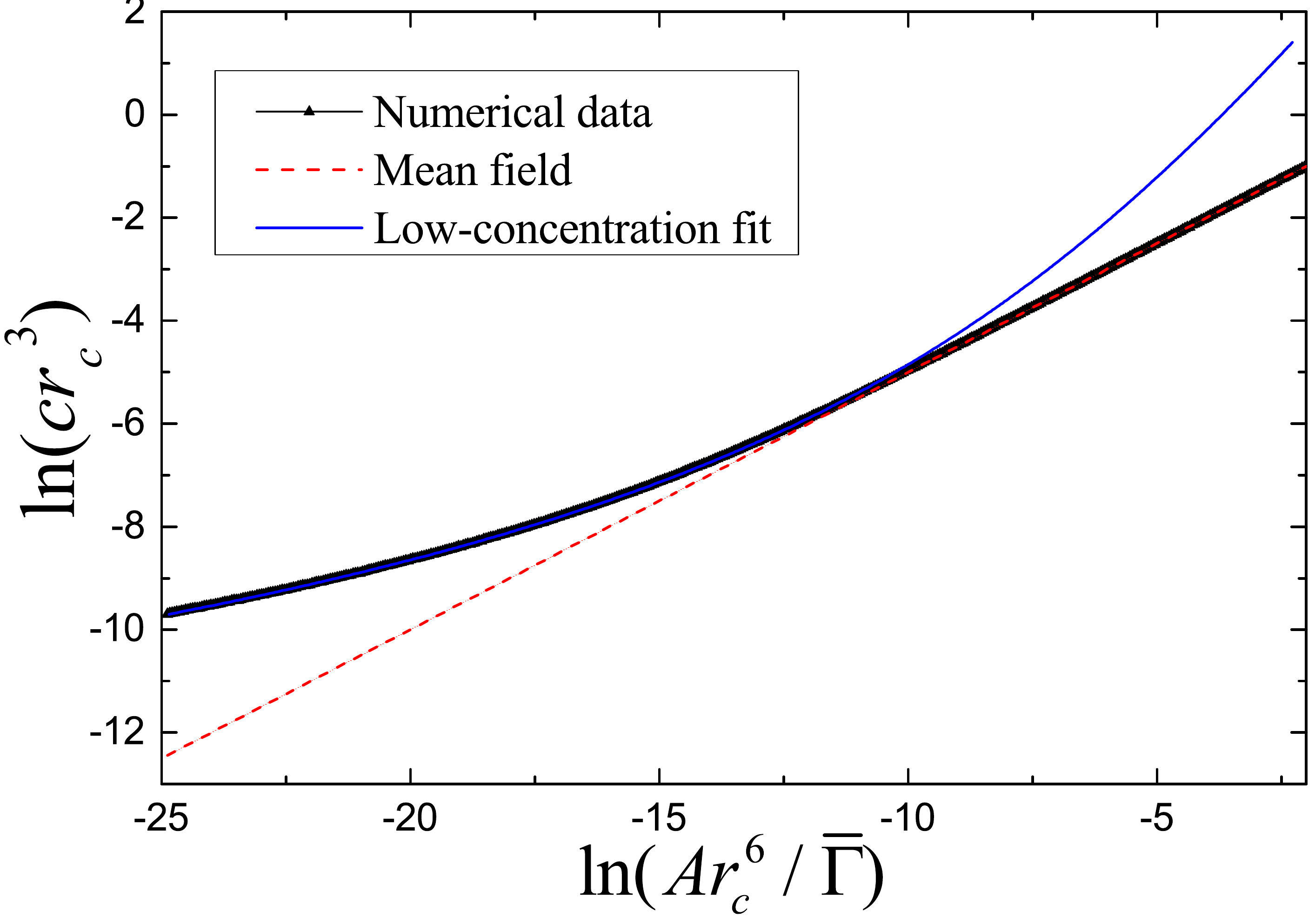}
\caption{\label{fig:LogFit} Numerical dependence of the quasiparticle concentration on the generation rate (black line), obtained from the simulation with spinless quasiparticles. The low-concentration fit is given by  $c=C_p/(4\pi r^3/3)$, where $C_p=0.605$ and $r$ is defined by Eq.~(9) in the main text, and the mean-field fit is given by $c = (A/\bar{\Gamma})^{1/2}$.}
\end{figure}

Starting with an empty box, typically 20000 simulation steps were enough to reach equilibrium, where the concentration $c = N/L^3$ stayed approximately constant. After equilibrium was reached, we determined the average concentration by averaging the particle number $N$ over the next 100000 simulation steps. The resulting $c$ vs $A$ graph is shown in Fig. \ref{fig:LogFit}. While at large injection rates the mean-field approximation provides a good fit, the concentration at small injection rates largely exceeds the mean-field estimation. The low-concentration part of the numerical data $c \lesssim 5\times 10^{-3} r_c^{-3}$, is well fitted by 
 $c=C_p/(4\pi r^3/3)$, where $C_p=0.605$ and $r$ is defined by Eq.~(9) in the main text
%Eq.~(10) in the main text.

\begin{figure}
\includegraphics[width = \linewidth]{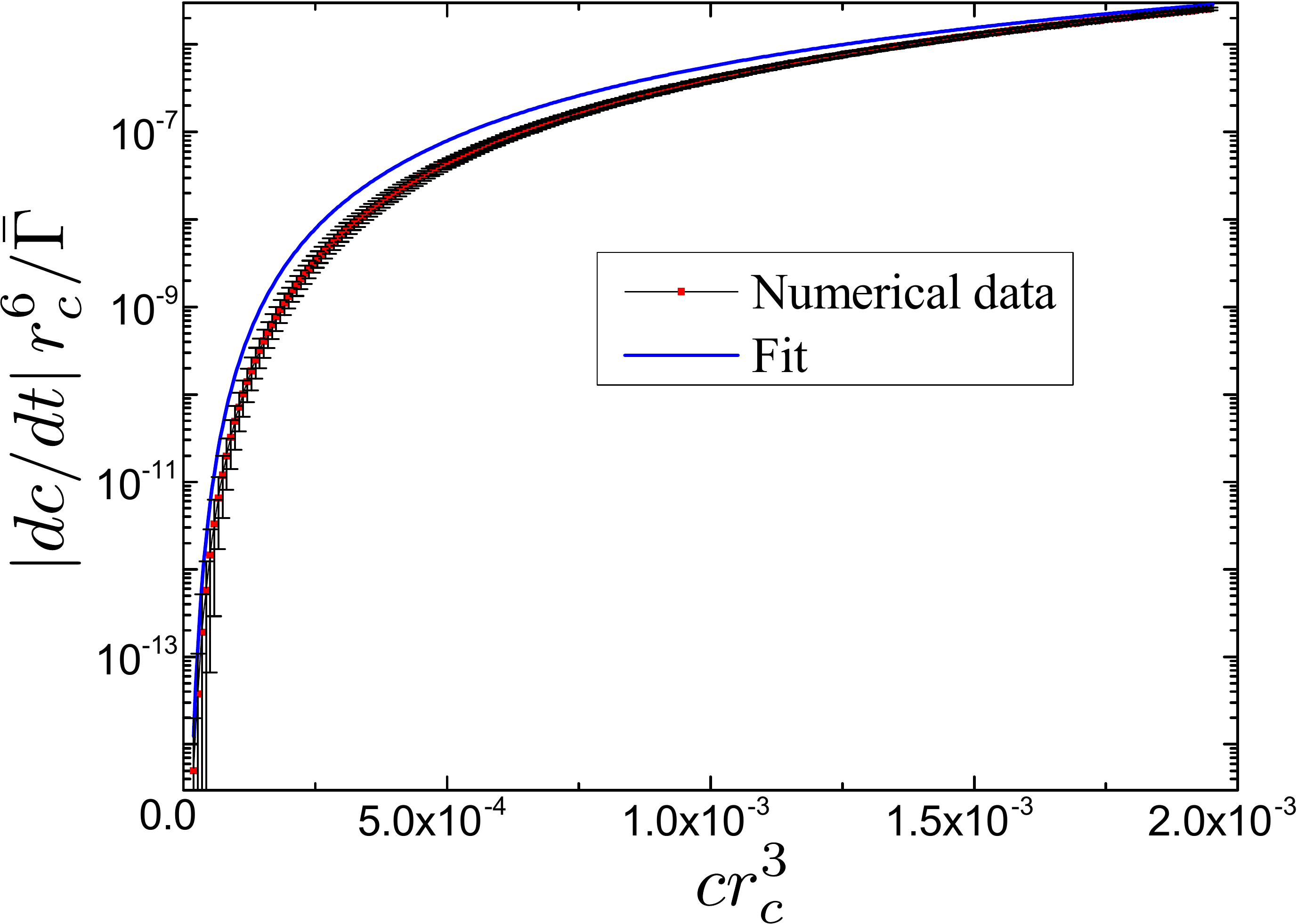}
\vspace{0.0cm}
\caption{\label{fig:Relaxation} The relaxation rate as a function of quasiparticle concentration in the absence of quasiparticle injection ($A = 0$). The error bars (black) indicate the standard deviation of $dc/dt$ when averaging over 1000 iterations of the relaxation process. The fit is given by  Eq.~(8) in the main tex.}
\end{figure}

In a similar way, we modeled the relaxation of quasiparticles in the absence of quasiparticle injection, $A=0$. To do so, we first let the the system reach equilibrium using a fairly large injection rate (we used $A r_c^6/\bar{\Gamma} = 6 \times 10^{-6}$, such that the equilibrium number of quasiparticles is $N \approx 1400$  for a box of size $L = 80r_c$). Then we monitored the evolution of the concentration after switching off the quasiparticle injection and considering only annihilation processes. Namely, at each step two quasiparticles are annihilated, and the relaxation  
rate $dc/dt$, which is given by
\begin{equation}
	\frac{dc}{dt} = L^{-3} \frac{dN}{dt} = -\frac{2 \Gamma_{\rm tot}}{L^3},
	\label{eq:dc/dt}
\end{equation}
is recorded. The recording of data was started once the number had decreased to $N = 1000$.
The whole relaxation process was repeated 1000 times to obtain averaged values for $dc/dt$. The resulting numerical data points together with the fit by  Eq.~(9) in the main text are shown in Fig. \ref{fig:Relaxation}. It can be seen that the fit qualitatively describes the numerical data, but it is nevertheless significantly worse than for the equilibrium concentration, Fig. \ref{fig:LogFit}. This may be explained by two factors. First, Eq.~(9) in the main text does not account for the spatial distribution of quasiparticles. This is expected to play a more important role when the system is far away  from equilibrium. Second, the relaxation simulation was done with a smaller number of quasiparticles in the box on average. Thus, the numerical data is of lower quality.

\section{Simulation of the bursting bubbles model with spin}
\label{sec:Bubbles}

Within the bursting bubbles model with spin, each quasiparticle is represented by a bubble with radius $r/2$ carrying a classical spin that is  either up or down. Only two overlapping bubbles with \emph{opposite} spins burst.

In our simulation, the system volume is a box with periodic boundary conditions and size $L \times L \times L$. We chose $L = 30r$, as it gave the best performance in terms of accuracy and simulation time. As a check, several runs with $L = 45r$ were done, yielding the same quasiparticle concentration in equilibrium. 

In each simulation step, either a quasiparticle with random spin is generated with probability $p_{\rm gen}=AL^3/(AL^3 +N \tau_{\rm sf}^{-1})$, or the spin of a random quasiparticle is flipped with probability $p_{\rm sf}=1- p_{\rm gen}= N \tau_{\rm sf}^{-1}/(AL^3+N\tau_{\rm sf}^{-1})$, 
%so that $p_{\rm gen} + p_{\rm sf} = 1$, 
where $\tau_{\rm sf}^{-1}$ is the  spin-flip rate. After this step, if the bubble of the added or flipped particle intersects with a bubble with opposite spin, the two bubbles ``burst'', and the corresponding particles are removed from the system.

\begin{figure}
\includegraphics[width = \linewidth]{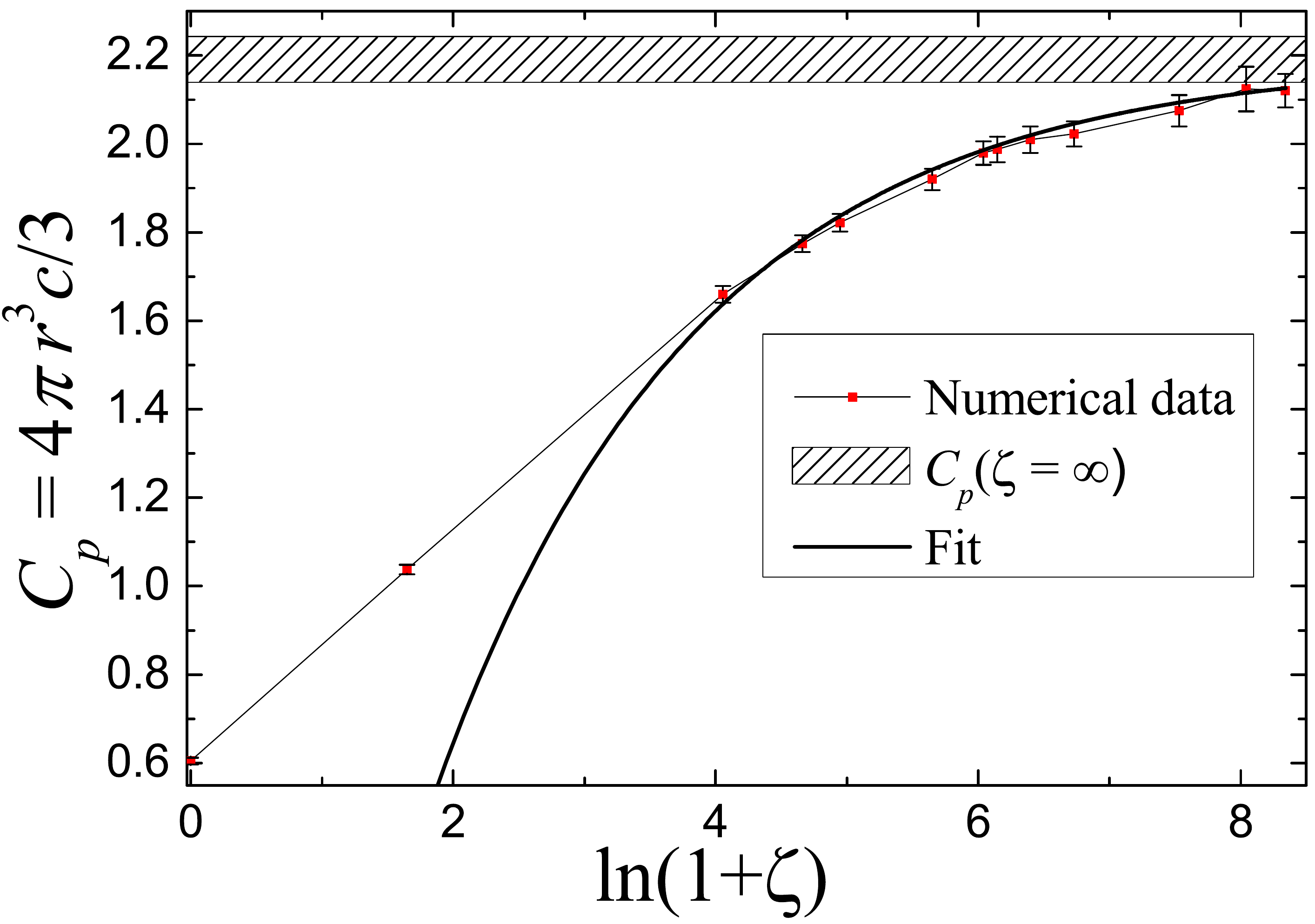}
\caption{\label{fig:cp} Concentration $c$ [in units of $3/(4\pi r^{3})$] as a function of the spin-flip time $\zeta$ (in units of $4\pi A r^3/3$) in the limit of instantaneous annihilation (bursting bubbles model). Here $r$ is the diameter of the bubbles. The variance is indicated by vertical bars. The hashed region indicates the concentration as well as its variance in the absence of spin flips, $\zeta=\infty$. The fit is given by Eq.~\eqref{eq:Cp_approx}.}
\end{figure}

\begin{figure}
\includegraphics[width = \linewidth]{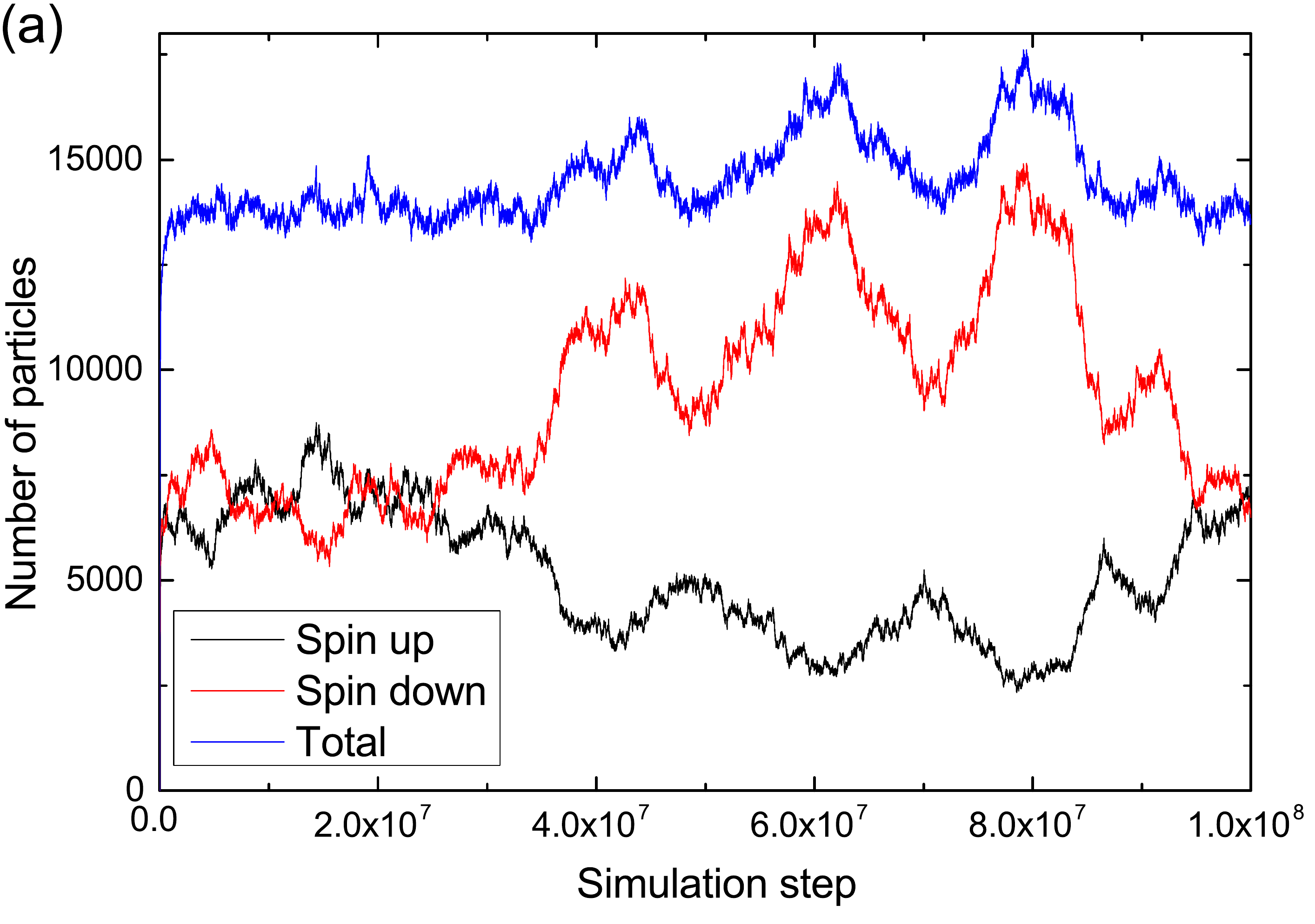}
\includegraphics[width = \linewidth]{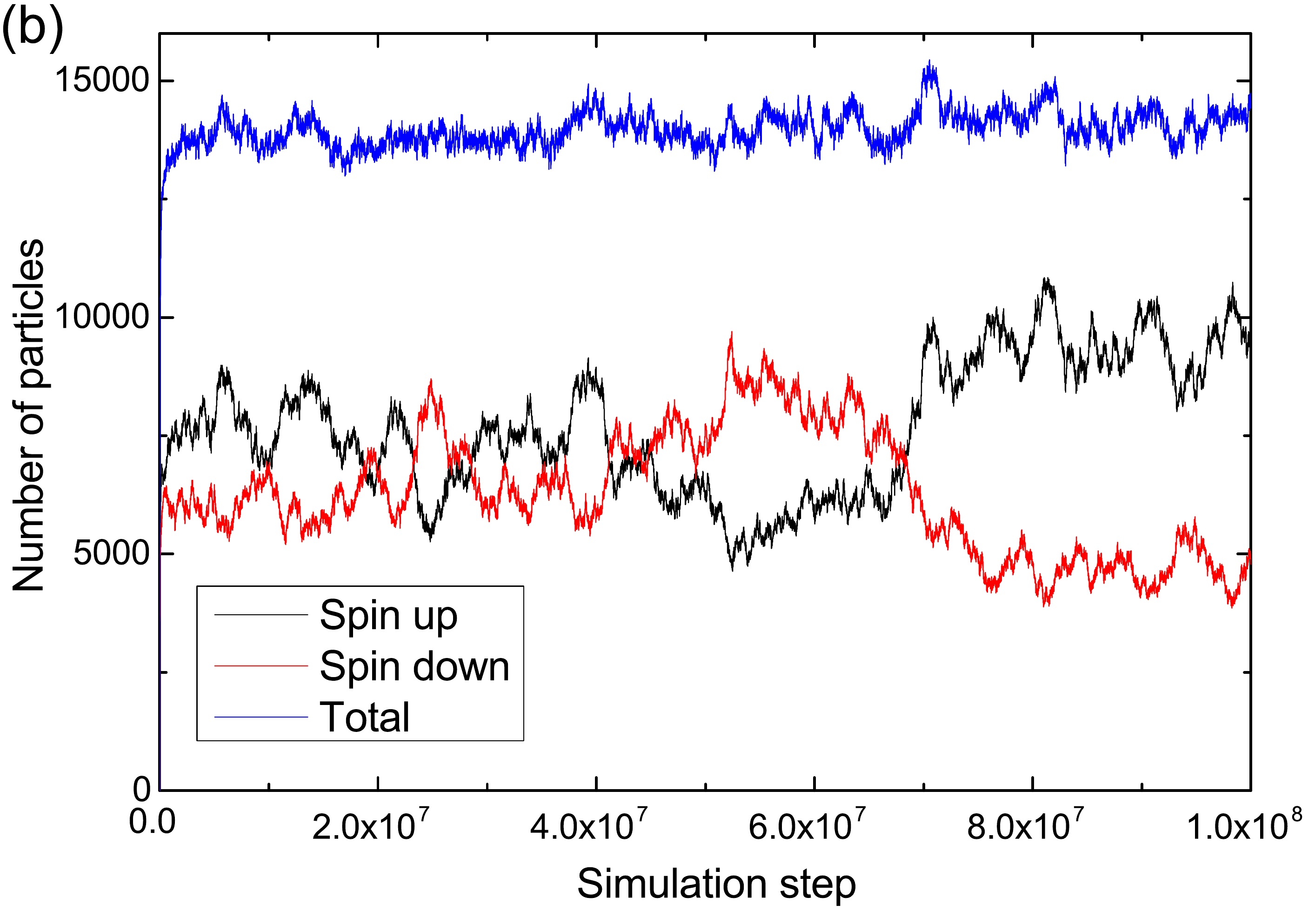}
\caption{\label{fig:Evolution} Number of particles vs simulation step in the bursting bubbles model without spin flips. The two graphs correspond to different realizations of the random process.}% In the simulation, the size of the box was $30 \times 30 \times 30$, bubble radius $r/2 = 0.5$.}
\end{figure}

In equilibrium, the dimensionless concentration (packing coefficient) $C_p = 4\pi cr^3/3$ is a function of the dimensionless spin-flip time $\zeta=4\pi A r^3\tau_{\rm sf}/3$ only. We varied this parameter in the range $\zeta = 0 - 4200$. The case $\zeta = 0$ corresponds to an effectively spinless system, such that the value $C_p(\zeta = 0) = 0.605$ is recovered. In addition, a simulation without spin flips was performed, which corresponds to $\zeta = \infty$. In that case, we obtained $C_p(\infty) = 2.19 \pm 0.05$. The dependence of $C_p$ on $\zeta$  is shown in Fig.~\ref{fig:cp}. It can be seen that the packing coefficient reaches the value $C_p(\infty)$ rather slowly with growing $\zeta$. In particular, we found that $C_p(\zeta)$ at $\zeta \gg 1$ is well approximated by
\begin{equation}
	C_p(\zeta) = C_p(\infty) - \frac{4.2}{\sqrt{\zeta}}.
	\label{eq:Cp_approx}
\end{equation}
The time evolution of the concentration in the absence of spin flips, $\zeta=\infty$, was also recorded. Typical graphs of the particle number vs simulation step dependence are shown in Fig.~\ref{fig:Evolution}. It can be seen that the system exhibits large fluctuations on a time scale of the order of $10^7$ simulation steps. Note that this requires averaging over many simulation steps when determining $C_p(\infty)$. Specifically, the average of the concentration was taken over $5\times 10^7$ steps.

\begin{figure}
  \includegraphics[width = \linewidth]{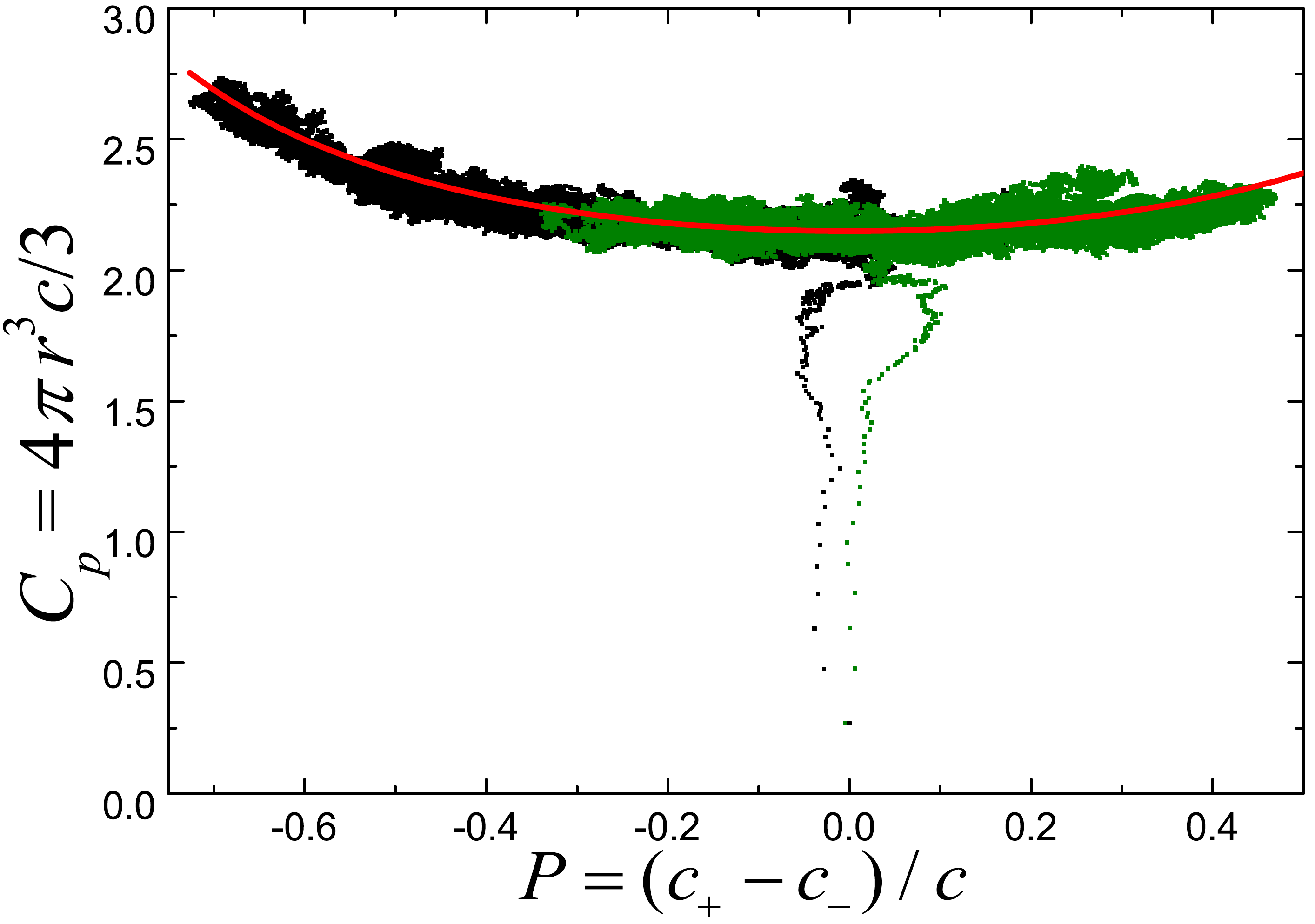}
	\caption{\label{fig:Polarization} Concentration vs spin polarization obtained from the two runs of the bursting bubbles simulation shown in Fig.~\ref{fig:Evolution}. The red curve is the fit by Eq.~\eqref{eq:c_fit}.}
\end{figure}

Another interesting feature that we observed is the strong correlation between the concentration and the instantaneous polarization of the system, $P = (c_+ - c_-)/c$, where $c_+$ and $c_-$ are the concentrations of spin-up and spin-down quasiparticles, respectively. Certainly, $P = 0$ on  average, however, the polarization exhibits strong fluctuations that are accompanied by fluctuations of the concentration. 
 Using the system evolution data, we obtained the $c$ vs $P$ graph shown in Fig.~\ref{fig:Polarization}. To qualitatively explain the $c$ vsvs $P$ dependence, we use the following argumentation. When a new quasiparticle is added to the system, there are four possibilities: in a sphere with radius $r$ encircling the quasiparticle, there might be (i) no quasiparticles, (ii) only quasiparticles of the same spin, (iii) only quasiparticles of the opposite spin, or (iv) quasiparticles of both spins present. In cases (i) and (ii) the total number of quasiparticles increases by 1, whereas in cases (iii) and (iv) it decreases by 1. Since for incoming quasiparticles both directions of spin  are equally probable, the probabilities of (ii) and (iii) are always the same. Thus, the stationary concentration is achieved if the probabilities for (i) and (iv) are equal as well, $p_{\rm(i)}=p_{\rm(iv)}$. To estimate these probabilities, we make the rough assumption that the positions of all quasiparticles do not correlate. This assumptions allows us to determine the probability to find no spin-up/spin-down quasiparticle in a sphere with radius $r$: $p_{\pm} = \exp[-4\pi r^3 c_{\pm}/3]$. Then, $p_{\rm(i)} = p_+ p_-$ and $p_{\rm(iv)} = (1-p_+)(1-p_-)$. Hence, in equilibrium $p_+ + p_- = 1$, or
\begin{equation}
	\frac{2\pi}{3} cr^3 = \ln \left[ 2 \cosh \left( \frac{2\pi}{3} Pc r^3 \right) \right].
	\label{eq:c_rough}
\end{equation}
At $P=0$, Eq.~\eqref{eq:c_rough} would yield $C_p(\infty)=\ln 2$, which is significantly smaller than the value obtained from the simulation. In reality, the quasiparticle positions correlate, which allows one to achieve higher concentrations. A better fit of the $c$ vs $P$ numerical data is obtained, if the concentration is increased by a factor of $1.55$:
\begin{equation}
	\frac{2\pi}{3} \frac{cr^3}{1.55} = \ln \left[ 2 \cosh \left( \frac{2\pi}{3} \frac{Pc r^3}{1.55} \right) \right].
	\label{eq:c_fit}
\end{equation}
This corresponds to the value of $C_p = 2.19$ found in the previous simulation.

\begin{thebibliography}{99}
\bibitem{note1}
The same parameter fixes the large energy range, $\varepsilon_T\ll\varepsilon \ll \varepsilon_T N_T^2$, in which the tail states are determined by optimal~\cite{LarkinOvchinnikov}, rather than mesoscopic~\cite{MeyerSimons}, fluctuations, according to Ref.~\cite{SkvortsovFeigelman}.
\bibitem{note3}
With $\nu_0\approx 2\times 10^{10}\,$eV$^{-1}\mu$m$^{-3}$, $\Delta\approx 200\,\mu$eV, and $D\approx 20\,{\rm cm}^2{\rm s}^{-1}$, we find $g\approx 10^4$. We also use $\varepsilon_T\lesssim\varepsilon_g\lesssim \Delta$ to estimate $N_T$.
\end{thebibliography}
\end{document}